# Self-Powered Triboelectric Sensing System for Gait-Based Physiological and Psychological Assessment in Track and Field

Tiehuai Liang[1], Dongyuan Wei[2*], Qi Zhang[3]

[1]College of Physical Education and Health, Guangxi Normal University, Guilin, 541000, China;
[2]Department of Sports, Guilin University of Electronic Technology, Guilin, 541000, China;
[3]School of Electronic Engineering and Automation, Guilin University of Electronic Technology, Guilin, 541000, China

*Corresponding author: weidy@guet.edu.cn.

***Abstract:*** Wearable sensors have become essential components in smart sports systems for real-time monitoring of athletic performance, physiological conditions, and psychological states. In this work, a sodium alginate/gelatin-based triboelectric nanogenerator (SG-TENG) was developed for mechanical energy harvesting and real-time monitoring in track and field applications. The SA/Gelatin composite film exhibits excellent transparency, flexibility, and homogeneous morphology, enabling stable triboelectric performance. The SG-TENG delivers a high output with a peak open-circuit voltage ($V_{OC}$) of 156.6 V, short-circuit current ($I_{SC}$) of 46.9 μA, and transferred charge ($Q_{SC}$) of 139.6 nC, achieving a maximum power of 13.5 mW under optimal load. Its output performance is strongly dependent on mechanical parameters such as frequency, force, displacement, and contact area. In addition, the device shows effective energy storage capability by charging capacitors under various conditions. Integrated into a running shoe, the SG-TENG enables self-powered gait monitoring and accurately distinguishes physical activities including walking, running, and jumping. Furthermore, it demonstrates the ability to infer psychological and physiological states from gait dynamics, highlighting its potential for battery-free, multifunctional sensing in sports performance and health monitoring.

## 1. Introduction

In the era of smart sports and big data-driven athletic analytics, the integration of real-time sensing technologies with intelligent performance evaluation has become a central focus in modern sports science [1-4]. Advanced data acquisition systems are increasingly being applied to monitor athletes' movement, physical condition, and training load, aiming to optimize performance, reduce injury risk, and support personalized training strategies [5-7]. Track and field sports, including sprinting, hurdling, long jump, and throwing, demand precise coordination of posture, stable psychological engagement, and sustained physiological output [8]. During high-intensity training or competition, improper posture or movement imbalance can impair athletic performance and increase the likelihood of injury [9].

Simultaneously, psychological states—such as tension, anxiety, or reduced focus—and physiological conditions—such as muscular fatigue or cardiopulmonary stress—can subtly influence gait patterns and movement dynamics [10, 11]. To meet the complex demands of modern sports science, there is a growing need for wearable sensing technologies capable of capturing multi-dimensional information that reflects not only biomechanical motion but also the athlete's mental and physical status [12-17]. While existing commercial wearable sensors, such as piezoresistive and capacitive sensors, have achieved significant progress in tracking muscle activity and physiological signals, they are often limited by their dependence on external power sources, relatively rigid designs, and a focus on single-parameter measurements [18-20]. Conventional sensing systems, limited by power dependence, rigid configurations, and single-modality detection, are poorly suited for long-term, multidimensional monitoring in dynamic environments like track and field training. These challenges underscore the urgent demand for next-generation wearable sensors that are lightweight, energy-independent, and capable of capturing nuanced, real-time signals related to athletic movement, psychological state, and physiological performance. Developing such technologies will be crucial for enabling more intelligent, personalized, and effective training systems in competitive sports environments.

Triboelectric nanogenerator (TENG) has emerged as a promising technology for mechanical-to-electrical energy conversion, particularly suitable for harvesting low-frequency and low-amplitude mechanical energy in ambient environments [21–37]. This mechanism is fundamentally based on the synergistic effect of contact electrification and electrostatic induction, wherein mechanical motion between two dissimilar materials leads to the generation and redistribution of surface charges. When these surfaces undergo periodic contact and subsequent separation, a time-varying potential difference is established between the electrodes, driving electron flow through an external circuit and thus producing alternating electrical output signals [38–43]. Depending on the nature and direction of the relative motion involved, TENGs can be classified into several working modes, including but not limited to vertical contact–separation mode, lateral sliding mode, single-electrode mode, and freestanding triboelectric-layer mode [44-53]. Each configuration offers distinct advantages in terms of structural integration, adaptability, and energy output characteristics, enabling their application across a wide range of self-powered sensing and energy harvesting scenarios. These diverse working modes endow TENGs with excellent structural flexibility, high sensitivity, and broad environmental adaptability. Owing to their strong responsiveness to low-frequency, intermittent, and small-amplitude mechanical stimuli, TENGs have demonstrated significant potential in harvesting ambient mechanical energy sources, such as wind and water wave energy, as well as bio-mechanical energy from human activities including walking, limb movement, and joint flexion [54-56]. In the domain of wearable electronics, TENGs have been extensively explored for harvesting human kinetic energy to drive self-powered systems, thereby enhancing device autonomy and operational sustainability [57, 58]. Moreover, by capturing bio-mechanical signals such as plantar pressure, gait rhythm, and joint motion, TENG-based sensors can facilitate real-time assessment of physical status, fatigue levels, and even psychological states during exercise [59]. Sodium alginate (SA), a natural, biodegradable, and biocompatible polysaccharide, has been widely explored as a triboelectric material due to its abundant functional groups and excellent film-forming capability [60]. Recent studies have demonstrated that SA-based composite films can be engineered with various nanomaterials, such as MXene, $TiO_2$, graphene oxide, and PVA, to enhance output performance, flexibility, and environmental stability [61-66]. SA has also been used to develop humidity-resistant, washable, and even antibacterial triboelectric nanogenerators for diverse applications including energy harvesting, motion sensing, and physiological monitoring. These advancements highlight the potential of SA as a versatile platform for next-generation self-powered systems.

In this study, we developed a triboelectric nanogenerator (SG-TENG) based on a sodium alginate (SA) and gelatin composite for harvesting bio-mechanical energy and enabling psychological and physiological monitoring in track and field training scenarios. The SA/gelatin film exhibits favorable properties including high optical transparency, excellent mechanical flexibility, smooth surface morphology, and robust inter-molecular hydrogen bonding. These features contribute to the formation of a homogeneous amorphous network that supports consistent triboelectric behavior, making the material well-suited for wearable electronics. Under periodic contact–separation conditions, the SG-TENG delivers strong electrical performance, achieving a peak open-circuit voltage ($V_{OC}$) of 156.6 V, a short-circuit current ($I_{SC}$) of 46.9 μA, and a maximum transferred charge ($Q_{SC}$) of 139.6 nC. Additionally, a peak output power of 13.5 mW is obtained at the matched load, confirming the device's high energy conversion efficiency and practical potential in self-powered systems. Notably, the device's electrical output is highly sensitive to variations in mechanical parameters such as contact force, frequency, displacement, and area, where increased values consistently lead to enhanced signal strength. Besides, the device demonstrates effective energy storage capability by charging capacitors under varying conditions, highlighting its potential for self-powered sensing and portable energy supply. The SG-TENG, integrated into a running shoe, enables real-time, self-powered monitoring of gait patterns, effectively distinguishing various physical activities such as walking, running, and jumping. Beyond motion classification, it can also infer psychological and physiological states through subtle variations in gait dynamics, demonstrating its potential for comprehensive, battery-free sensing in sports and health monitoring.

## 2. Experiments

*2.1. Preparation of SA/Gelatin composite film.*

The SA/Gelatin composite film was prepared through a solution casting method involving stepwise dissolution, blending, and vacuum drying, as schematically illustrated in Figure 1a–d. In the first step, gelatin powder was added to deionized water and stirred continuously at an elevated temperature of 80 °C (Fig. 1(a)). This thermal treatment facilitated the transition of gelatin from a disordered coil conformation into an ordered gel-like network. The heat-induced denaturation partially disrupted the intramolecular hydrogen bonds in gelatin, allowing polypeptide chains to rearrange and form an interconnected physical network, which is essential for film-forming capability and mechanical stability. In parallel, sodium alginate (SA) powder was dispersed in deionized water under vigorous stirring at room temperature to form a stable colloidal suspension (Fig. 1(b)). The ionic polysaccharide chains of SA readily dissolved in water, yielding a homogeneous and low-viscosity dispersion that serves as the complementary matrix component in the final composite. Subsequently, the thermally treated gelatin solution and SA dispersion were mixed together in appropriate weight ratios to form a uniform SA/Gelatin solution (Fig. 1(c)). This mixture contained both gel-like gelatin networks and ionically active SA chains, enabling strong intermolecular interactions via hydrogen bonding and electrostatic forces. The resulting hybrid solution was poured into a clean mold and subjected to vacuum drying at 60 °C for 8 hours. During this step, water was gradually removed under reduced pressure, leading to densification and film formation. The intermolecular interactions between gelatin and SA were further enhanced during the drying process, contributing to the structural integrity of the resulting film. As shown in Fig. 1(d) a freestanding SA/Gelatin composite film was successfully obtained. The film exhibited good flexibility, transparency, and uniformity, making it suitable for subsequent applications in TENGs and flexible electronic devices.

*2.2. Preparation of SG-TENG device.*

The SG-TENG device was assembled in a vertical contact–separation configuration, as illustrated in Fig. 1(e). The device structure consists of two symmetrical friction layers and electrode layers, with the SA/Gelatin composite film serving as the triboelectric active layer. A copper foil (serving as the bottom electrode) was first laminated onto a flexible Kapton substrate using a thin adhesive layer to ensure firm contact and mechanical stability. A layer of PVC film was then attached to the copper as the counter triboelectric surface. The SA/Gelatin film, previously prepared and vacuum dried, was carefully laminated onto a second copper/Kapton layer to form the complementary friction layer. All layers were aligned and stacked using adhesive bonding to ensure uniform contact during mechanical actuation. The final SG-TENG structure features a sandwich configuration: Kapton/Copper/PVC on one side and Kapton/Copper/SA-Gelatin on the other. The entire assembly was compact, flexible, and well-suited for periodic contact–separation motion under external force. The electrical output was measured across an external load resistance connected between the two copper electrodes. Fig. 1(f) presents a photographic comparison between the fabricated SA/Gelatin film and a commercial PTFE film used in the assembly of SG-TENG devices.

*2.3. Characterization and measurements.*

The electrical output performances of the TENG, including $V_{OC}$, $I_{SC}$, and $Q_{SC}$, were measured using a Keithley 6514 electrometer. A programmable linear motor was employed to simulate periodic contact–separation motion. A force sensor was integrated beneath the device to monitor the applied pressure in real-time, and a feedback loop dynamically adjusted the motor displacement to ensure controlled mechanical excitation. The entire setup was housed within a temperature and humidity-controlled chamber ($25 \pm 1$ °C, $50 \pm 5\%$ RH) to minimize environmental variation and ensure repeatability.

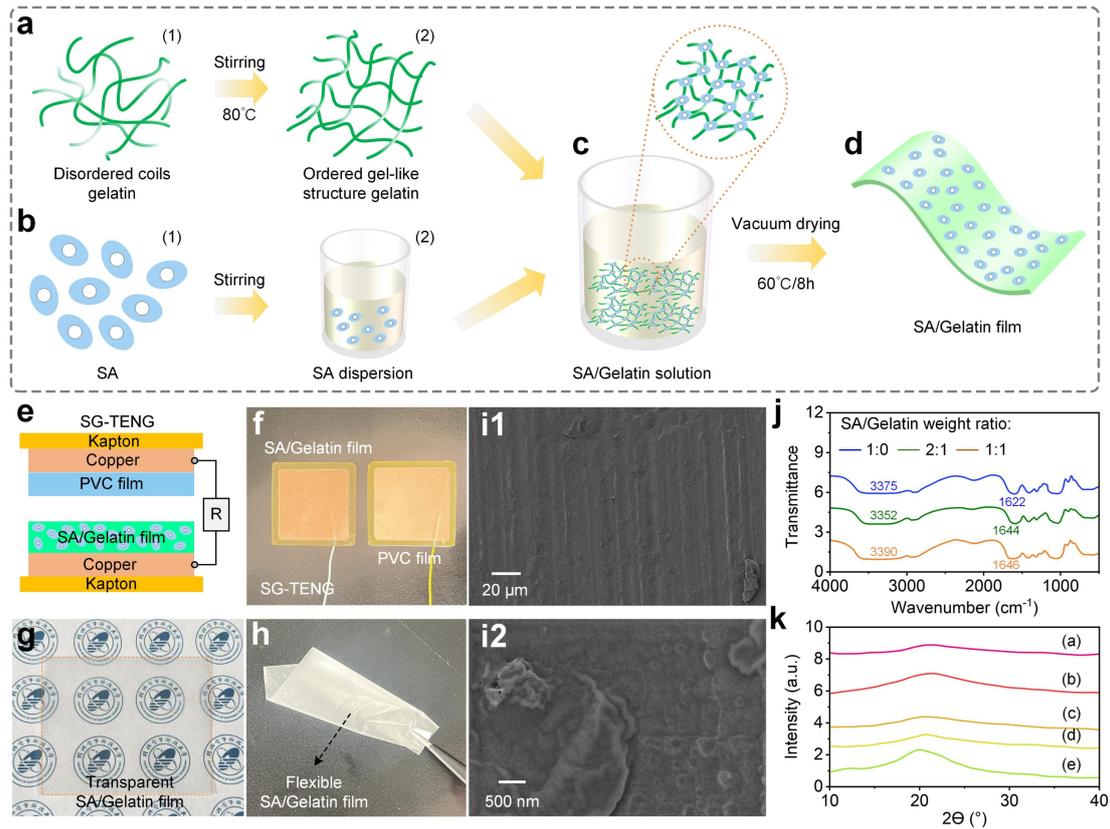

Fig. 1. Schematic illustration of the fabrication process for the SA/Gelatin composite film: (a) Gelatin solution formation by heating and stirring; (b) Preparation of SA dispersion; (c) Mixing of SA and gelatin to form a homogeneous solution; (d) Vacuum drying to obtain the SA/Gelatin film. (e) Schematic structure of the SG-TENG device based on SA/Gelatin and PVC friction layers. (f) Optical image of SA/Gelatin and PTFE films used for comparison. (g) Transparent SA/Gelatin film placed on patterned paper. (h) Flexible and bendable SA/Gelatin film. SEM images showing the surface (i1) and cross-sectional morphology (i2) of the SA/Gelatin film. (j) FTIR spectra of SA/Gelatin films with different weight ratios. (k) XRD patterns of SA/Gelatin films with varying compositions.

## 3. Results and discussion

*3.1. Characterization of SA/Gelatin composite film.*

As shown in Fig. 1(g), the film exhibits excellent transparency, allowing clear visibility of underlying patterns, which confirms its suitability for optical or wearable applications. Fig. 1(h) highlights the high flexibility of the film, which can be easily bent and folded without visible cracking or delamination, indicating good mechanical robustness. The surface and cross-sectional morphologies were further examined by scanning electron microscopy (SEM). The surface image (Fig. (i1)) reveals a smooth and dense microstructure with minimal defects, while the high-magnification image (Fig. (i2)) displays a compact internal structure with uniformly distributed components. These features confirm the homogeneous blending of SA and gelatin, as well as the structural integrity of the composite film, which is essential for stable triboelectric performance in SG-TENG devices. Fig. 1(j) illustrates the FTIR spectra of composite films with different SA to gelatin weight ratios, highlighting the molecular interactions between SA and gelatin. The absorption peak at 1645 cm$^{-1}$ corresponds to the Amide I band of gelatin, confirming its incorporation into the film matrix. A shift in the broad O-H stretching band from 3400 cm$^{-1}$ to lower wavenumbers (e.g., 3375 cm$^{-1}$) is observed as gelatin is added, indicating the formation of inter-molecular hydrogen bonds between SA and gelatin. These spectral changes reflect strong physical interactions between the two bio-polymers. The absence of any new peaks further confirms that the films are physically cross-linked rather than chemically bonded. Fig. 1(k) displays the X-ray diffraction (XRD) patterns of composite films prepared with different SA to gelatin weight ratios. All samples exhibit broad diffraction peaks in the range of 2θ = 20°–22°, characteristic of amorphous polymeric structures. The broadness of the peaks indicates a low degree of crystallinity, which is typical for bio-polymer-based films. As the gelatin content increases, the intensity and sharpness of the diffraction peaks change slightly, suggesting alterations in the internal molecular packing. Notably, the film with a moderate SA/gelatin ratio shows a more pronounced peak, indicating better molecular ordering or inter-molecular interaction at that specific ratio. These results confirm that the structural arrangement of the polymer chains is influenced by the relative composition of SA and gelatin, while maintaining an overall amorphous nature.

*3.2. The working mechanism of SG-TENG device.*

The operational principle of the SG-TENG follows a vertical contact–separation mechanism, as schematically depicted in Fig. 2. In the initial resting state (Fig. 2(a1)), the triboelectric layers—comprising a polyvinyl chloride (PVC) film and a sodium alginate/gelatin (SA/Gelatin) composite film—remain physically separated, resulting in no electric potential across the external electrodes. When mechanical pressure is applied (Fig. 2(a2)), the two surfaces are brought into intimate contact, initiating triboelectric charge generation due to their differing electron affinities. Specifically, electrons are transferred from the more tribo-positive SA/Gelatin layer to the more tribo-negative PVC layer, leading to the accumulation of opposite charges on each surface. As the layers reach full contact (Fig. 2(a3)), electrostatic equilibrium is established at the interface, and the system momentarily ceases to drive charge flow through the circuit. Upon release of the applied force, the layers separate (Fig. 2(a4)), disrupting the balance and inducing a potential gradient between the electrodes. This newly formed potential drives electrons through the external circuit to neutralize the charge imbalance, thereby generating a transient electrical current. This cyclic process of contact and separation continually modulates the electrostatic field, resulting in alternating charge transfer and a corresponding alternating current (AC) output. Furthermore, the polarity of the output signal reverses with each compression–release cycle, consistent with the inherent bidirectional nature of the triboelectric effect. The excellent dielectric properties and efficient charge storage capability of the SA/Gelatin film play a vital role in enhancing the charge separation and maximizing the electrical output of the device.

*3.3. The output performance of SG-TENG device.*

The electrical output characteristics of the SG-TENG under periodic contact–separation motion were systematically investigated, as shown in Fig. 2(b-f). Fig. 2(b) presents the $V_{OC}$ profile over time, where the SG-TENG delivers a peak $V_{OC}$ of approximately 156.6 V. The voltage waveform exhibits excellent periodicity and stability, confirming the reliability of the device under continuous operation. The corresponding $I_{SC}$ is shown in Fig. 2(c), reaching a maximum value of 46.9 μA with sharp and symmetric current peaks, indicating fast charge transfer and low internal impedance during separation.

The $Q_{SC}$ output was measured using the charge integration mode of a Keithley 6514 electrometer, as illustrated in Fig. 2(d). The peak charge reached up to 139.6 nC, demonstrating effective charge accumulation enabled by the strong triboelectric interaction between the SA/Gelatin and PVC layers. To assess the electrical adaptability of the SG-TENG under different external conditions, a range of load resistances from 1 MΩ to 1000 MΩ was introduced, with the corresponding voltage and current responses presented in Fig. 2(e). As the resistance value increased, a rising trend in voltage and a simultaneous decline in current output were observed, consistent with Ohmic behavior. The variation of output power with respect to load resistance is illustrated in Fig. 2(f), where the SG-TENG achieved a peak power output of 13.5 mW at an optimal resistance of approximately 20 MΩ. This result indicates effective impedance matching, which facilitates maximum energy transfer. Overall, the analysis confirms the SG-TENG's excellent power delivery capability and highlights its strong potential for real-world energy harvesting applications.

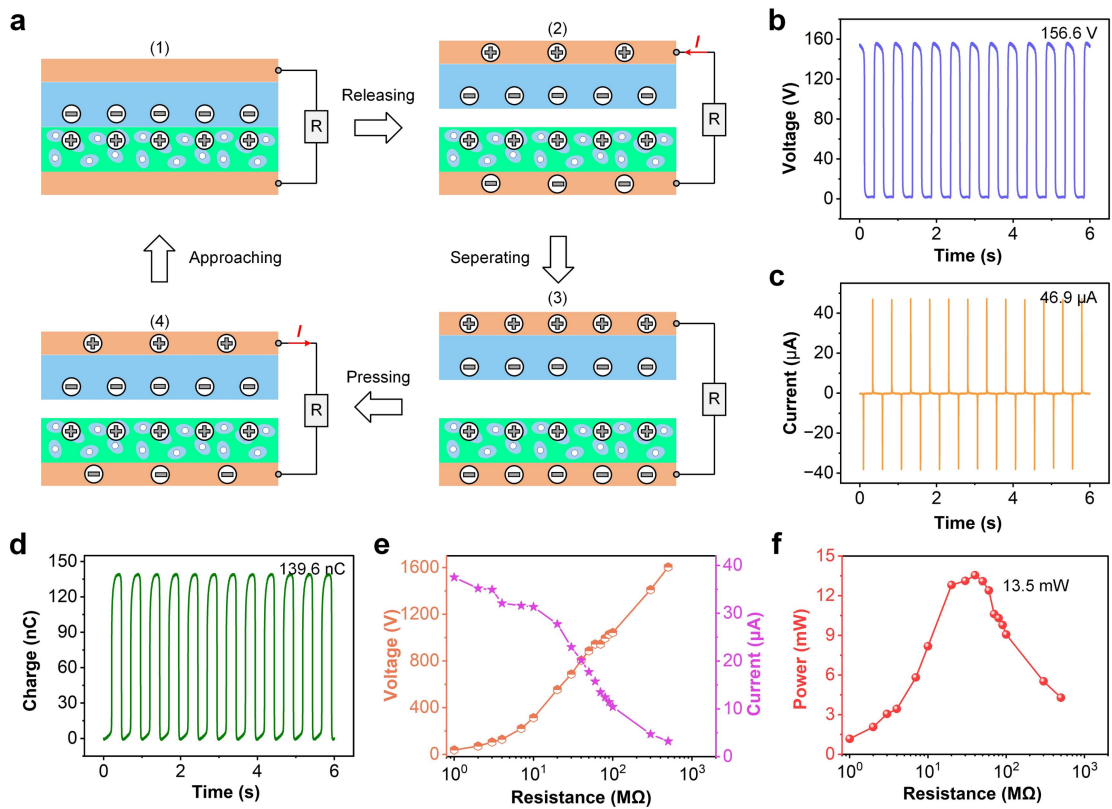

Fig. 2. (a1–a4) Working mechanism of the SG-TENG in a contact–separation mode, illustrating charge transfer and current flow during periodic mechanical motion. (b) $V_{OC}$, (c) $I_{SC}$, and (d) $Q_{SC}$ output under periodic contact–separation cycles. (e) Dependence of output voltage and current on external load resistance. (f) Output power as a function of load resistance, showing a maximum power of 13.5 mW.

The frequency-dependent output performance of the SG-TENG was evaluated under a constant applied normal force of 5 N and a fixed vertical displacement of 3 mm. These conditions were chosen to ensure consistent contact between friction layers while minimizing structural deformation. Fig. 3(a–c) depict the $V_{OC}$, $I_{SC}$, and $Q_{SC}$ under varying contact–separation frequencies ranging from 2 Hz to 6 Hz at a constant applied force. As the frequency increases, the output voltage (Fig. 3(a)) remains relatively stable, reaching up to ~157.5 V at 6 Hz. This indicates the robustness of the SG-TENG device in maintaining voltage generation across a wide dynamic range. In contrast, the $I_{SC}$ (Fig. 3(b)) increases significantly with frequency, rising from ~15.3 μA at 2 Hz to ~25.9 μA at 6 Hz. This enhancement can be attributed to the increased rate of charge separation and transfer per unit time. The $Q_{SC}$ (Fig. 3(c)) also shows a slight increase with frequency, stabilizing around 76.1 nC at 6 Hz, reflecting efficient and repeatable charge generation under dynamic cycling. Fig. 3(d–f) present the electrical output behavior of the SG-TENG under different applied normal forces, from 1 N to 8 N, at a fixed operating frequency. As shown in Fig. 3(d), the $V_{OC}$ increases monotonically with the applied force, reaching ~116.5 V at 8 N. This is due to the improved contact intimacy and effective contact area between the two triboelectric layers under higher pressure, which enhances surface charge density. Similarly, the current output (Fig. 3(e)) also increases proportionally with force, reaching a peak of ~17.8 μA at 8 N. The transferred charge (Fig. 3(f)) shows a corresponding increase, peaking at ~88.9 nC under 8 N force. These results confirm that both higher frequencies and larger applied forces are beneficial for boosting the electrical performance of SG-TENG, by enhancing the contact quality and the dynamic charge separation rate. Such tunability highlights the potential of SG-TENG for adaptable energy harvesting and sensing in varying mechanical environments. Although the maximum tested normal force in these controlled experiments is 8 N—lower than a typical adult's body weight—this value is sufficient to simulate the localized transient pressure acting on specific areas (e.g., heel) of the in-shoe sensor during gait. In practical applications, the distributed contact force is buffered by shoe materials and foot geometry, resulting in localized effective forces within the tested range.

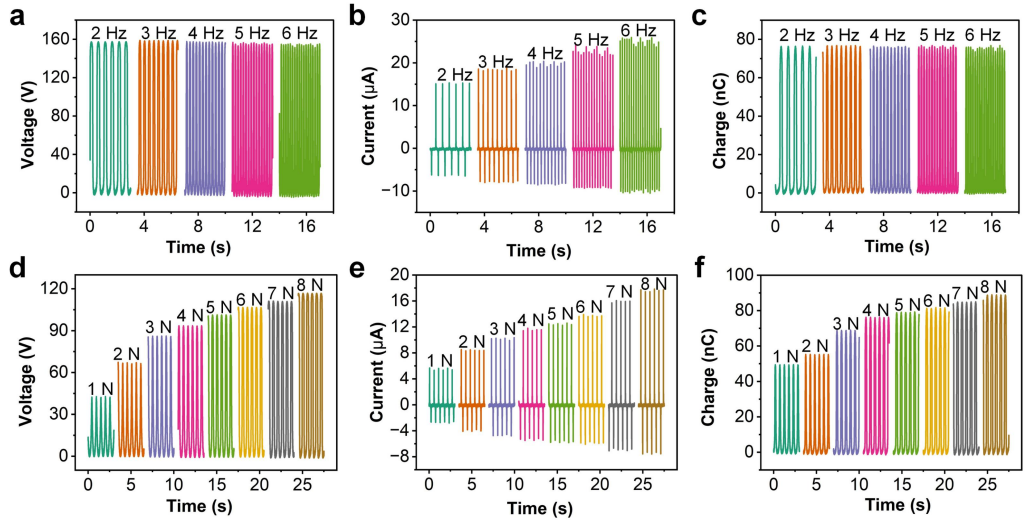

Fig. 3. Electrical output performance of the SG-TENG under different operating frequencies (2–6 Hz) at constant applied force: (a) $V_{OC}$, (b) $I_{SC}$, and (c) $Q_{SC}$. The current increases significantly with frequency due to accelerated charge transfer, while voltage and charge outputs remain relatively stable. Electrical outputs of the SG-TENG under varying applied forces (1–8 N) at a fixed frequency: (d) $V_{OC}$, (e) $I_{SC}$, and (f) $Q_{SC}$.

In this section, the term "contact distance" refers to the vertical displacement applied to the top triboelectric layer (SA/Gelatin side) during contact–separation cycles, controlled via the linear motor system. This displacement simulates the vertical motion that occurs in practical applications and directly influences the degree of mechanical contact and separation between the triboelectric layers. Fig. 4(a–c) show the variations in output voltage, current, and transferred charge as the vertical contact displacement increases from 1 mm to 5 mm. As shown in Fig. 4(a), the $V_{OC}$ increases from ~90.8 V to ~144.1 V with increasing displacement, primarily due to enhanced mechanical interaction and increased charge separation distance during contact–separation cycles. Fig. 4(b) shows a similar trend for $I_{SC}$, which increases from ~18.4 µA to ~32.1 µA, indicating accelerated charge transfer at higher displacements. Correspondingly, the transferred charge also rises with increasing displacement, reaching ~107.7 nC at 5 mm (Fig. 4(c)). These results confirm that larger displacement enhances effective contact and separation, improving the triboelectric output. The influence of contact area was further examined by systematically reducing the effective contact area from 80% to 20% (Fig. 4(d–f)), while maintaining constant mechanical excitation. As shown in Fig. 4(d), the $V_{OC}$ decreases proportionally with decreasing area, dropping from ~123.2 V at 80% to ~78.4 V at 20%. Similar reductions are observed in the $I_{SC}$ (Fig. 4(e)) and $Q_{SC}$ (Fig. 4(f)) outputs, indicating that effective triboelectric charge generation is strongly dependent on the contact area. The decrease in surface interactions leads to reduced charge density and limited energy conversion capability. The energy harvesting and storage capability of the SG-TENG was also investigated through a typical rectifier–capacitor circuit, as illustrated in Fig. 4(g). A full-wave bridge rectifier was used to convert the AC output to DC, which was then stored in external capacitors. The capacitor charging profiles under different capacitances are presented in Fig. 4(h). When charged using the SG-TENG operating at 5 Hz, a 1 µF capacitor was charged to ~16.5 V within 150 s, whereas a 3.2 µF capacitor reached ~13.4 V in the same time frame, indicating a slower voltage rise due to larger capacitance but higher energy storage capacity. Additionally, the effect of operation frequency on capacitor charging was explored. As shown in Fig. 4(i), a 1 µF capacitor charged at higher frequencies (3, 4, and 5 Hz) exhibited faster voltage growth, with the fastest rate observed at 5 Hz. This is attributed to increased charge generation per unit time at higher mechanical frequencies, enabling more efficient energy transfer into storage components. These results collectively demonstrate that both mechanical displacement and contact area significantly affect SG-TENG performance, and that the SG-TENG can effectively convert bio-mechanical energy into storable electrical energy through efficient rectification and capacitive storage.

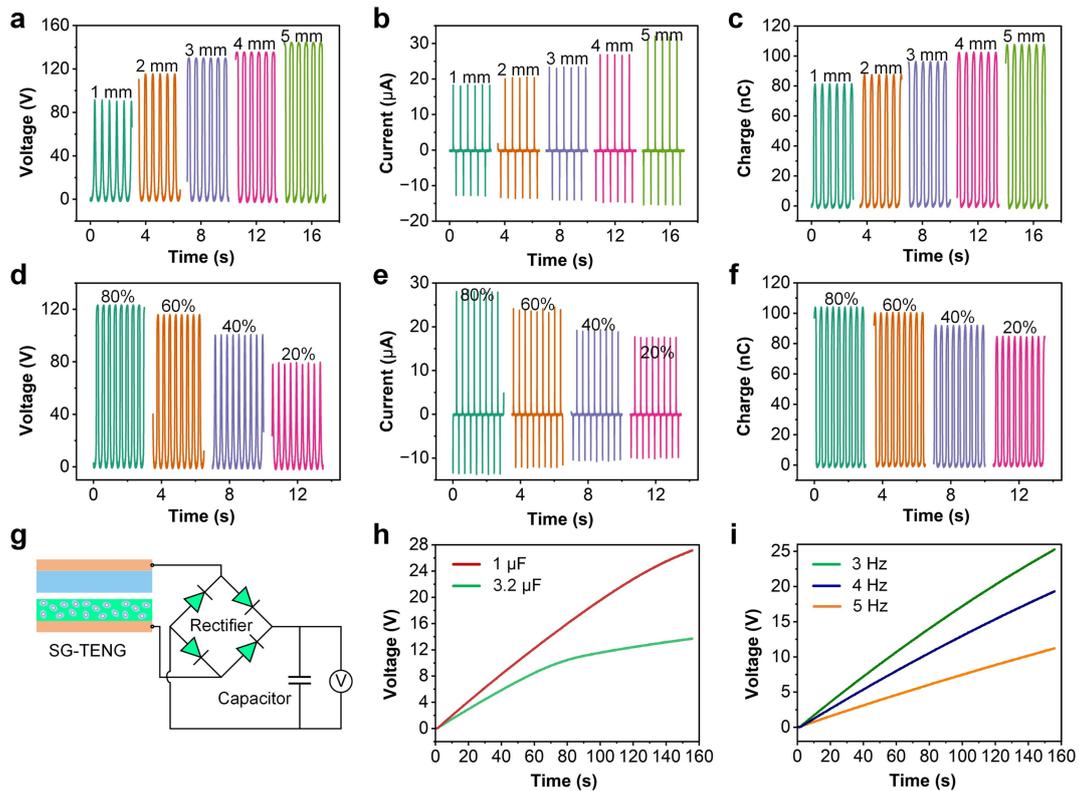

Fig. 4. Effects of external parameters on the electrical performance of the FC-TENG. (a) $V_{OC}$, (b) $I_{SC}$, and (c) $Q_{SC}$ of the FC-TENG at different operating frequencies (2–6 Hz). (d–f) Dependence of $V_{OC}$, $I_{SC}$, and $Q_{SC}$ on applied force (5–25 N). (g–i) Influence of separation distance (1–9 mm) on $V_{OC}$, $I_{SC}$, and $Q_{SC}$.

To explore the practical applicability of the SG-TENG in wearable motion sensing and health monitoring, a prototype system was developed by integrating the SG-TENG into a running shoe, as shown in Fig. 5(a). The triboelectric nanogenerator was placed at the heel position, where it could be compressed and released during foot-ground contact events. This setup enables real-time detection of gait patterns through the generated electrical signals. The working principle is illustrated in Fig. 5(b). During locomotion, the vertical motion of the foot induces periodic contact and separation between the SA/Gelatin layer and the PVC surface, generating current signals in response to mechanical motion. These signals are captured using a current meter connected in series with the external circuit. A typical single gait cycle produces a distinct current peak corresponding to the heel strike and lift-off phases, as shown in Fig. 5(c). Fig. 5(d) demonstrates the SG-TENG's ability to distinguish different athletic activities including walking, running, and jumping. Each motion type generates a unique signal pattern in terms of amplitude and frequency. Walking produces low-amplitude, low-frequency signals, while running generates higher frequency and larger amplitude signals due to increased impact and cadence. Jumping leads to sharp, high-current pulses caused by strong impact forces during landing. These distinct signatures indicate the capability of SG-TENG to perform motion classification in track and field sports. In addition to physical movement recognition, the SG-TENG was evaluated for psychological state monitoring, as shown in Fig. 5(e). Variations in gait related to emotional states, such as hesitation or tension, were reflected in the current signals. Neutral walking produced regular and symmetrical signals, whereas slowed and uncertain walking exhibited decreased amplitude and irregular timing, potentially indicating reduced confidence or anxiety. In contrast, hurried and tense walking produced high-frequency and asymmetrical signals, suggesting urgency or mental stress. These findings demonstrate that subtle psychological changes can be captured through gait dynamics, providing a non-invasive method for emotional state inference. Fig. 5(f) highlights the physiological monitoring capability of the system. When the user exhibited signs of fatigue or heavy breathing after sustained running, the current signal became irregular and showed fluctuations in amplitude and interval timing. Initially, energetic movement generated strong and stable signals. As the user became fatigued, the signal exhibited reduced periodicity and increased noise, reflecting labored or uncoordinated motion. The amplitude of the peaks decreased, suggesting weaker ground impact due to reduced step force or altered gait biomechanics. These patterns can be used to infer physiological states such as energy expenditure, endurance level, or onset of fatigue. Together, these results demonstrate the multifunctional potential of SG-TENG as a wearable sensing platform for comprehensive gait analysis. The ability to detect not only basic locomotion modes but also infer psychological and physiological variations from gait dynamics highlights its value in sports science, rehabilitation, and mental health monitoring. Importantly, the system is self-powered, eliminating the need for external energy sources, making it suitable for continuous, real-world use in athletic environments. While the current study does not employ advanced signal processing or machine learning algorithms, the observed variations in time-domain electrical signals—such as amplitude, frequency, and periodicity—provide intuitive indicators of user state. These variations correlate with physical manifestations of fatigue (e.g., reduced force, irregular gait), emotional states (e.g., hesitation, urgency), and activity type. The purpose of this work is to demonstrate that the SG-TENG can sensitively capture such gait-dependent signal features in a self-powered manner. Although these features are not yet quantitatively classified, they form the basis for future integration with lightweight processors and data-driven analysis for real-time monitoring. Future work will focus on embedding signal feature extraction and classification modules to enable practical deployment in sports and health applications.

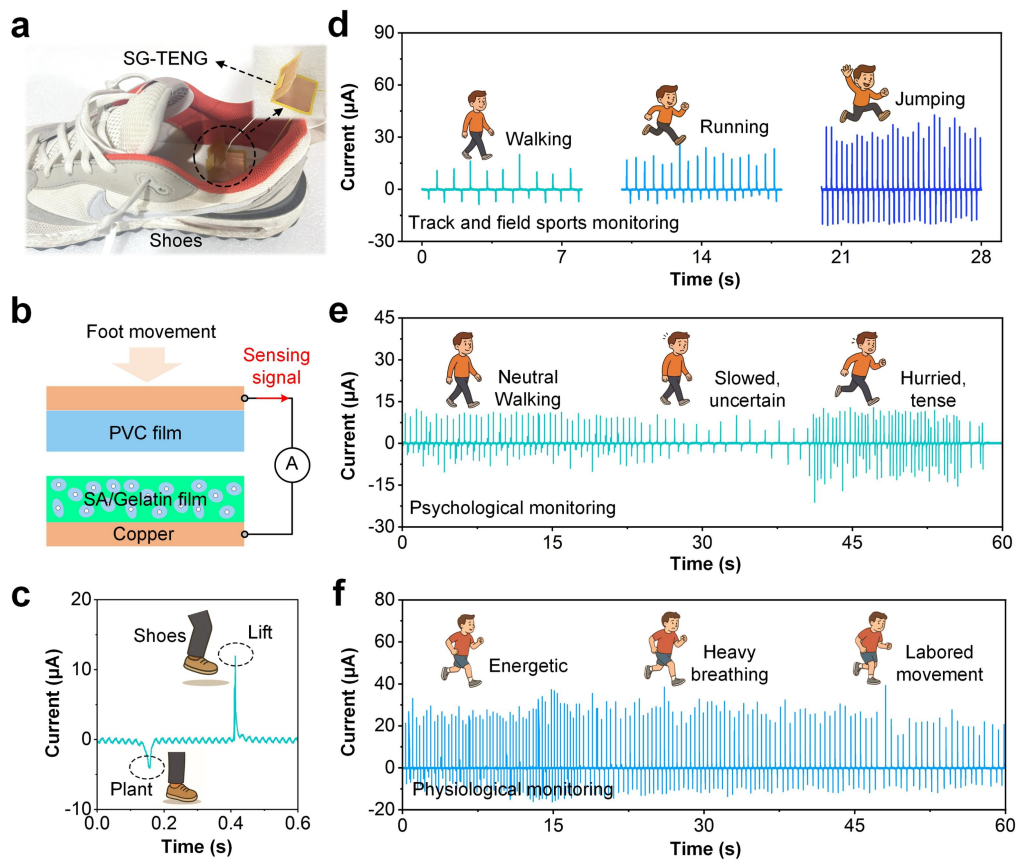

Fig. 5. (a) Photograph of the SG-TENG integrated into a running shoe for real-time gait sensing. (b) Working mechanism of the SG-TENG under foot motion, generating electrical signals via contact–separation. (c) Current response of a single gait cycle showing distinct peaks during heel strike and lift-off. (d) Current signals corresponding to different athletic motions (walking, running, jumping), demonstrating the capability for track and field activity monitoring. (e) Psychological state monitoring based on gait: variations in walking behavior reflect neutral, hesitant, or tense emotional states. (f) Physiological monitoring through fatigue analysis: signal irregularities correspond to energetic, heavy breathing, and labored movement stages.

## 4. Conclusions

In summary, we developed a SG-TENG capable of harvesting bio-mechanical energy and enabling real-time monitoring of physiological and psychological states in track and field sports. The SA/Gelatin composite film demonstrates excellent transparency, flexibility, and structural uniformity, ensuring stable triboelectric output for wearable applications. The SG-TENG achieves high electrical performance, with a peak $V_{OC}$ of 156.6 V, $I_{SC}$ of 46.9 μA, and $Q_{SC}$ of 139.6 nC, along with a maximum output power of 13.5 mW. Its output characteristics are tunable via mechanical parameters such as contact frequency, force, and area, and the device also exhibits reliable energy storage capability by charging capacitors under various conditions. When integrated into footwear, the SG-TENG enables self-powered gait monitoring, accurate activity classification, and the inference of mental and physical states from gait dynamics. This work demonstrates the potential of biodegradable, flexible materials in self-powered sensing systems and highlights the feasibility of integrating TENG-based devices into smart sports applications. In future research, combining SG-TENG with wireless transmission, data-driven analytics, and multi-sensor fusion strategies may further enhance its applicability in intelligent motion tracking, athlete health assessment, and early warning systems for sports-related fatigue or stress.

# 5. Data Availability Statements

Data available on request from the authors.